# Pressure induced superconductivity in $BaFe_2As_2$ single crystal

**Awadhesh Mani[*], Nilotpal Ghosh, S. Paulraj, A. Bharathi, and C. S. Sundar**

Materials Science Division,
Indira Gandhi Centre for Atomic Research
Kalpakkam -603102, Tamilnadu, INDIA

**Abstract**

The evolution of pressure induced superconductivity in single crystal as well as polycrystalline samples of $BaFe_2As_2$ has been investigated through temperature dependent electrical resistivity studies in 0-7 GPa pressure range. While the superconducting transition remains incomplete in polycrystalline sample, a clear pressure induced superconductivity with zero resistivity at the expense of magnetic transition, associated with spin density wave (SDW), is observed in the single crystal sample. The superconducting transition temperature ($T_C$) is seen to increase upto a moderate pressure of about ~1.5 GPa and decreases monotonically beyond this pressure. The SDW transition temperature $T_{SDW}$ decreases rapidly with increasing pressure and vanishes above ~1.5 GPa.

---

[*] Corresponding author's email: mani@igcar.gov.in

**INTRODUCTION**

The recent discovery of superconductivity in the iron pnictides [1-3] has evoked immense response from scientific community which has culminated in synthesis of several new superconductors pertaining to these classes of compounds. Amazingly, in a limited span of time four homologous series of the Fe- pnictides, namely, 1111 (ROFeAs with R = rare-earth; AeFFeAs with Ae = alkaline earth), 122 ($AeFe_2As_2$ with Ae = Ba, Ca, Sr), 111 (AFeAs with A=Li, Na) and 011 (FeSe) have been discovered. Proper doping by electrons or holes in these parent compounds has resulted in the maximum superconducting transition temperatures ($T_C$) of ~ 55 K for 1111 [4], ~ 38 K for 122 [5, 6], and 12-25 K for 111 [7, 8], and ~ 9-14 K for 011 [9], respectively. One distinguishing feature of these pnictide superconductors is the existence of spin density wave (SDW) in the parent compounds, which gives way to superconductivity with electron or hole doping [4-9]. Apart from the ample future scope of enhancing the $T_C$, as indicated in several doping studies, the occurrence of superconductivity in the presence of large concentration of the magnetic Fe in these layered compounds has provided an avenue to investigate the interplay of magnetism and superconductivity which may help in unraveling a long standing mystery of high temperature superconductivity.

High pressure techniques provide valuable tools for altering and understanding of the electronic, structural and ground state properties of materials without introducing any chemical complexity. Several such studies carried out on above classes of FeAs – compounds have revealed the following general trends [10]: (1) Pressure tends to decrease the SDW transition temperature in the undoped or slightly doped compounds. (2) $T_C$ increases with pressure for underdoped FeAs-pnictides, remains approximately constant for optimal doping, and decreases linearly in the overdoped range. (3) Pressure induces superconductivity in the parent LaOFeAs and AeFe$_2$As$_2$ at the cost of SDW transition. While the above general trends can be inferred from high pressure studies on these compounds, there are some uncertainties. For example, the high pressure magnetization studies performed by Alireza *et al*. [11] on the single crystals of BaFe$_2$As$_2$ had revealed a pressure induced bulk superconductivity with a maximum T$_C$ of 29 K above a critical pressure of 2.8 GPa. However, the high pressure electrical resistivity studies carried out by Fukazawa *et al*.[12] on polycrystalline samples indicated incomplete superconducting transition at about 20-30K above 2 GPa without any signature of zero resistivity. On the other hand, high pressure resistivity studies on single crystal by Torikachvili *et al*. [13] did not reveal any signature of the resistive superconducting transition.

We have carried out temperature dependent electrical resistivity studies on single crystal as well as polycrystalline samples of BaFe$_2$As$_2$. Our results show a clear pressure induced superconductivity with zero resistivity in single crystal with higher T$_C$ (35.4 K) at moderate pressure of ~1.5 GPa. Further, the polycrystalline sample exhibit incomplete superconducting transition under pressure with non-zero resistive background persisting upto the lowest measured temperature of 4.2 K.

**EXPERIMENTAL DETAILS**

Single crystals of BaFe$_2$As$_2$ have been grown by employing self-flux growth procedure in which excess FeAs is used as a flux [14]. FeAs was prepared first by mixing together in stoichiometric quantities of Fe and As powders followed by heat treatment at 600

$^{0}$C for 8 hours. Appropriate proportion of Ba has been weighed and kept in alumina crucibles (5 ml) along with FeAs in glove box. The crucible is then vacuum sealed in quartz ampoule, and heated up to 1100 $^{o}$C for 27 hrs and subjected by a programmed cooling at the rate of 5$^{o}$C /min in a box furnace. At the end, black, shiny, crystals have been observed embedded in solidified melt which were mechanically retrieved. The average size of the crystals is found to be 2mm x 1.5mm x 0.3mm. Formation of single crystal was affirmed by recording and analyzing the Laue patterns [15]. The preparation and characterization of polycrystalline sample are described in reference [6]. High pressure resistivity measurements as a function of temperature on a single crystal and powdered polycrystalline samples were carried out in an opposed anvil pressure locked cell. Steatite was used as pressure transmitting medium and pyrophyllite washers as gasket. Internal pressure of the cell was pre-calibrated by measuring the $T_C$ of Pb with respect to applied load prior to mounting the samples. The subsequent high pressure measurements on $BaFe_2As_2$ samples have been carried out by maintaining experimental conditions similar to that used during pressure calibration to minimize the error in pressure measurements. However, there can be an error of ~0.3 GPa in the reported pressure. Further details about the sample assembly and measurements can be found in reference [16,17].

**RESULTS AND DISCUSSION**

In Fig.1 we compare the temperature dependent normalized resistivity ($\rho/\rho_{300K}$) of polycrystalline as well as single crystal samples of $BaFe_2As_2$ recorded at ambient pressure. It is evident that both the samples exhibit metallic behavior with positive temperature coefficient of resistivity. The well known SDW anomaly [18] is clearly captured by the sharp change in the slopes of resistivity curves of the both the samples. The SDW transition temperatures ($T_{SDW}$) determined from the first derivative of the temperature dependent resistance as shown in the insets of Fig. 1 (a) & (b) for the respective sample are found to be 139 K, which is well within the reported values of 138-140 K for single crystal samples [14, 18].

Fig. 2 shows the evolution of the electrical resistance of polycrystalline powder $BaFe_2As_2$ sample as a function of temperature and pressure. A careful inspection of this figure shows a broad hump indicative of SDW transition near about $T_{SDW} \sim 102$ K at 1.6 GPa as marked by arrow. Here, $T_{SDW}$ is determined from the peak position of derivative curve shown in the inset of Fig 2. This feature cannot be discerned at higher pressures indicating vanishing of SDW transition beyond 1.6 GPa. At further lower temperatures, a sudden dip in resistance indicating onset of superconductivity between $T_C$ of ~ 25- 30 K for 1.6 - 4 GPa pressure respectively is observed. Here, $T_C$ is defined as the intersection point of extended tangents drawn in the two temperature ranges as depicted in one of the curve Fig. 2. Similar to what was seen in the previous studies [12], the resistance remains non-zero at all the pressures in present measurements also. It should be mentioned that incomplete resistive superconducting transition has also been reported in $SrFe_2As_2$ system [19] under pressure. There could be several reasons for the incomplete superconducting transition, namely, small superconducting fraction, poor grain contact [20] and anisotropy etc.

In Fig. 3 we present temperature and pressure dependent electrical resistivity $\rho(T,P)$ of single crystal of $BaFe_2As_2$ in 4-300K temperature and 0.8 – 7.2 GPa pressure range. These curves have been offset for clarity. A careful inspection of $\rho(T,P)$ curves in Fig. 3 reveals the following important features of the pressure effects: First, the signature of superconductivity with onset $T_C$ of 31.8 K is seen at pressure of 0.8 GPa, however, zero resistivity is not registered upto the lowest measured temperature of 4.2 K. Secondly, a complete superconducting transition with vanishing resistivity is seen at ~1.4 GPa pressure with $T_C$ of 35.4 K. It is to be remarked that in previous studies [11,12] the pressure induced superconductivity was seen above the critical pressure of 2.5 GPa and the maximum $T_C$ obtained was ~ 29 K in $BaFe_2As_2$ system. On the other hand, in the present study we find the largest pressure induced $T_C$ ( 35.4 K) at ~1.5 GPa. Third, the SDW transition temperature $T_{SDW}$ (marked by downward arrows and determined from the derivative plot) is observed to decrease rapidly with increasing pressure from 135 K at 0.8 GPa to 105 K at 1.4 GPa, which ultimately vanishes beyond this pressure. It should be mentioned that in polycrystalline sample also $T_{SDW}$ vanishes above similar magnitude of pressure P >1.6 GPa. It is noted that there is distinct change in the overall shape of the normal state ρ(T,P) curves beyond the pressure of 1.4 GPa above which SDW vanishes. For example at pressures P≤1.4 GPa, ρ(T,P) exhibits concave curvature below $T_{SDW}$ and transforms to convex curvature above $T_{SDW}$ resembling the resistivity curve of $MgCNi_3$ [21]. However, for P> 1.4 GPa, ρ(T,P) shows concave curvature through out like the ρ(T) of $MgB_2$ [22]. While SDW features are still seen, the appearance of complete superconductivity at P ~1.4 GPa indicates the coexistence of magnetism with superconductivity. Similar coexistence of SDW with superconductivity had been reported in $(Ba_{1-x}K_x)Fe_2As_2$ (x≤0.2) system for lower concentration of K from neutron diffraction and Mossbaur spectroscopy studies [23], and the SDW was found to disappear for higher concentration of K. It means that the superconductivity in $BaFe_2As_2$ appears at the brink of vanishing of the SDW in both the cases, namely, K-doping as well as under external pressure.

In order to gain more physical insight on the resistivity behaviour, we have fitted the normal state resistivity of $BaFe_2As_2$ at all the pressures using the following expression [24, 25]

$$\rho(T) = A + BT^m \quad \text{-----(1)}$$

In Eq. 1, the constant A represents the contribution to the resistivity from electron - impurity scattering and the second term accounts for contribution to resistivity from electron-electron interaction for m=2 and the anti-ferromagnetic spin fluctuation for m=1.5 [25]. The fits to experimental data in normal state are indicated in Fig 3. It is seen that the experimental data for pressure P ≤1.4 GPa fit to Eq. 1 for m = 1.5±0.07 in temperature range of 150-300 K *i.e.* above $T_{SDW}$. However, $\rho(T,P)$ fits the above expression in entire temperature range (40-300K) above superconducting transition for P≥2.0 GPa (a pressure beyond which spin density wave vanishes). From the $T^{3/2}$ dependence of resistivity, it may be inferred that the spin fluctuation dominates the normal state resistivity of this system. Parameter B is, thus, indicative of the strength of spin fluctuation. Parameter B increases systematically with pressure from $1.0 \times 10^{-5}$ at 0 GPa to $7.8 \times 10^{-4}$ m$\Omega$-cm/K$^2$ at 7.2 GPa, which may indicate strengthening of spin fluctuation with increasing pressure.

The variation of $T_{SDW}$, $T_C$ and fit parameter B of single crystal sample as a function of pressure is shown in Fig. 4. $T_{SDW}$ is seen to decrease drastically and vanish beyond P>1.4 GPa. On the other hand, $T_C$ shows a non-monotonic variation with pressure, *i.e.*, $T_C$ increases abruptly to 32.8 K at 0.8 GPa, reaches to a maximum of 35.4 K at 1.4 GPa and then decreases monotonically with further increase in pressure. It is worth mentioning that maximum of $T_C$ lies at the border of $T_{SDW}$ instability (cf. Fig. 4 a & b). This suggests that magnetic instability may be playing some role in the occurrence of superconductivity. The link between spin fluctuation and superconductivity in this system is also suggested in other studies [24, 26]. It is known that [27, 28] SDW instability in parent compounds arise due to the nesting of Fermi surface connecting the hole and electron Fermi surfaces by a commensurate vector $q=(\pi,\pi)$. Disruption of such a nesting condition by hole (electron) doping or pressure suppresses SDW order giving a way to the superconductivity. If the spin fluctuations are involved in the process of occurrence of superconductivity, then the increase of parameter B with pressure (cf Fig 4c) may suggest strengthening of superconductivity by increasing the coupling strength. However, $T_C$ decreases with pressure beyond ~1.4 GPa, while B increases through out. This decrease in $T_C$ may arise due to the reduction in DOS at Fermi energy $E_F$ with pressure. Detailed band structures calculations will be helpful in this regard in understanding of the manifestation of Fermi surface and the evolution DOS under pressure in this system. In effect, the non-monotonic variation of $T_C$ with pressure can arise due to the competing effect of increase of coupling strength, as indicated by B, and the reduction in DOS at $E_F$.

**SUMMARY**

In summary, we have measured temperature and pressure dependent electrical resistivity of single crystal as well as polycrystalline $BaFe_2As_2$ samples. Pressure induced superconductivity is observed in both the samples around ~1.5 GPa pressure. While superconducting transition remains incomplete in polycrystal, a full fledged superconductivity is observed in single crystal. $T_C$ exhibits an initial increase followed by decrease at higher pressures, while $T_{SDW}$ decrease rapidly and vanishes beyond 2 GPa pressure. Interestingly, superconductivity and magnetism coexist at moderate pressure of 1.4 GPa. The dominace of $T^{3/2}$ term in $\rho(T,P)$ data indicates the link between spin fluctuation and superconductivity in $BaFe_2As_2$. The non-monotonic variation of $T_C$ with pressure is suggestive of the competing effect of the increase of coupling strength and the deacrese in DOS at $E_F$..

**Acknowdgements:** Authors acknowledge R. M. Sarguna, D. Sornadurai and V. S. Sastry for the Laue and XRD characterization of single and poly crystalline samples.

# FIGURES

**Fig. 1 (a)-(b):** Variation of normalized resistivity of polycrystalline and single crystal samples respectively. Insets in both the panels show the derivative plot of resistivity to identify the SDW transition temperature $T_{SDW}$.

**Fig. 2:** Temperature and pressure dependent resistance plot of the polycrystalline $BaFe_2As_2$. The downward arrow shows the position of $T_{SDW}$ as determined from the first derivative plot shown in the inset of this figure.

**Fig. 3:** Temperature and pressure dependent resistance curves of the single crystal $BaFe_2As_2$ samples in various layers along with the fit as described in the text. The zero of each layer is indicated by horizontal thick line. Downward arrows indicate the positions of $T_{SDW}$ at various pressure P≥1.4 GPa.

**Fig. 4 (a)-(c):** Variation of $T_C$, $T_{SDW}$ and fit parameter B respectively as a function of pressure for single crystal sample.

Fig. 1

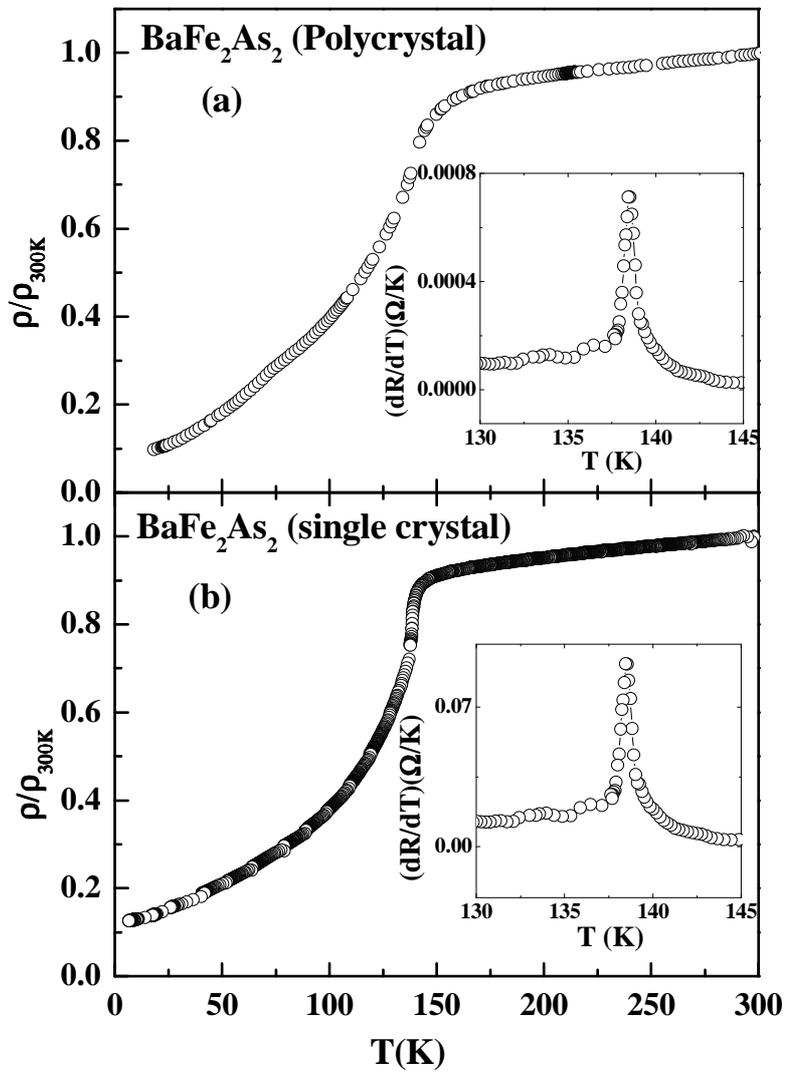

**Fig 2.**

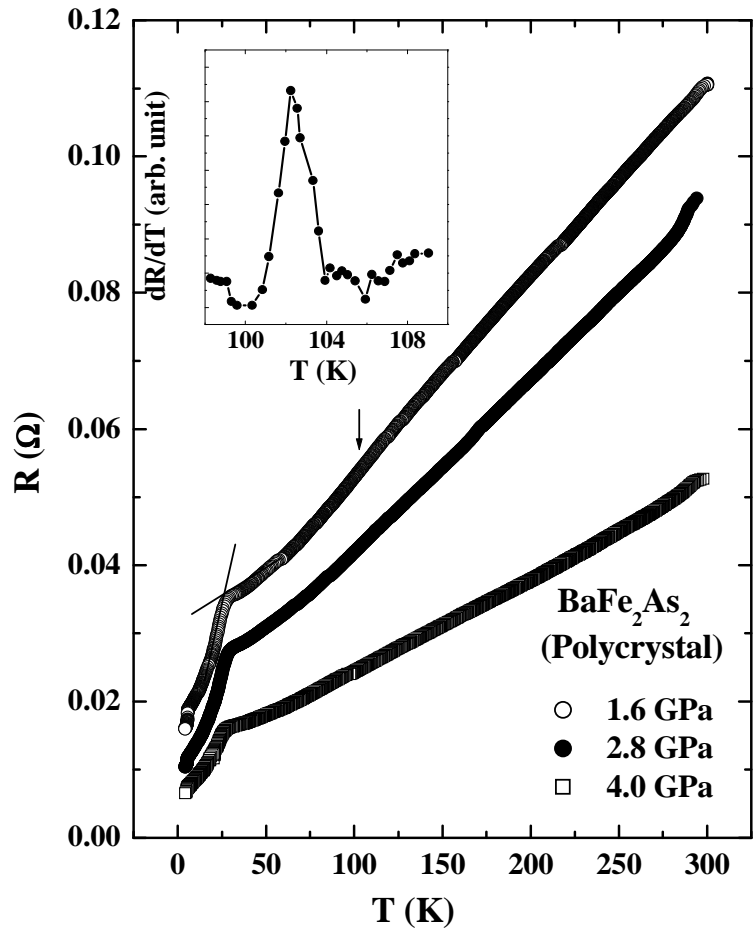

**Fig. 3**

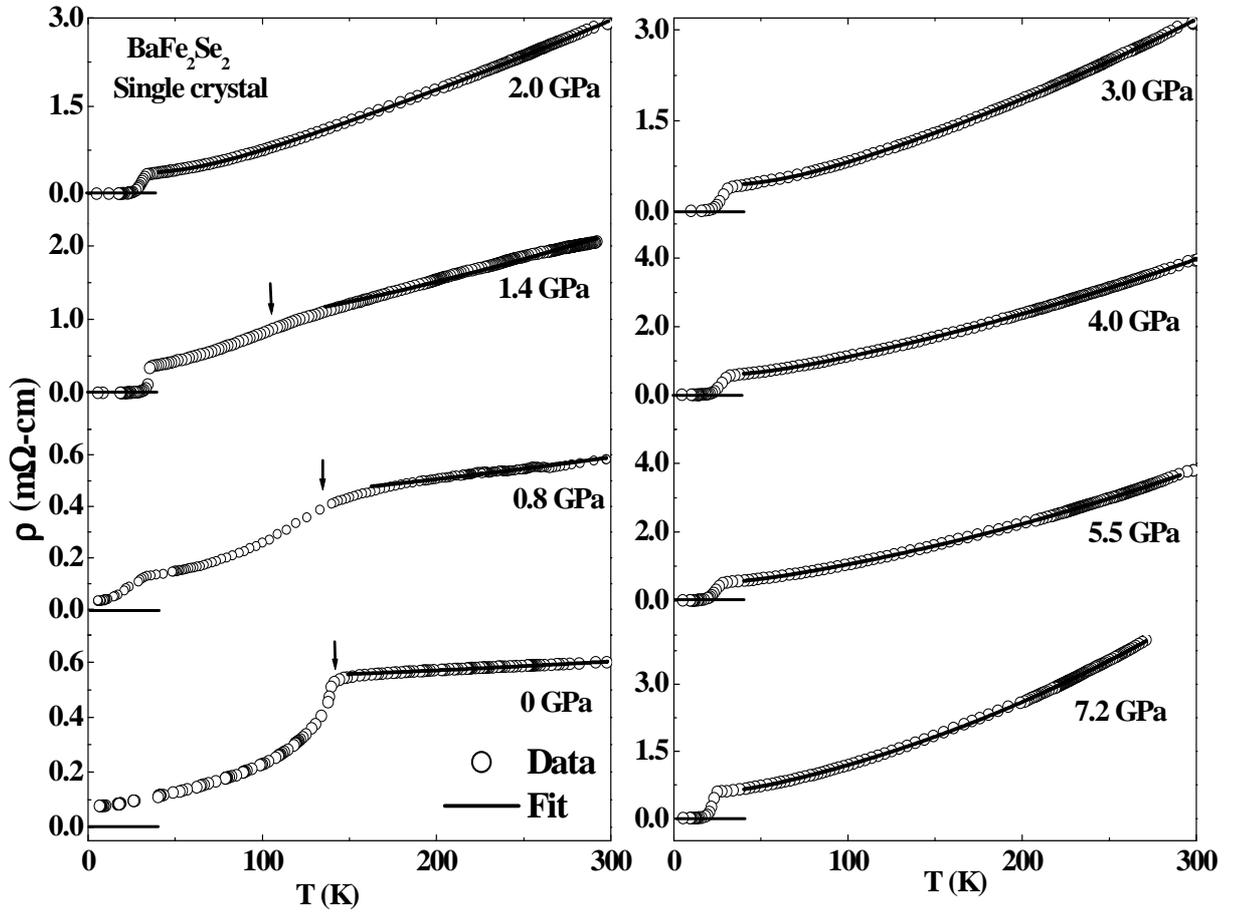

Fig. 4

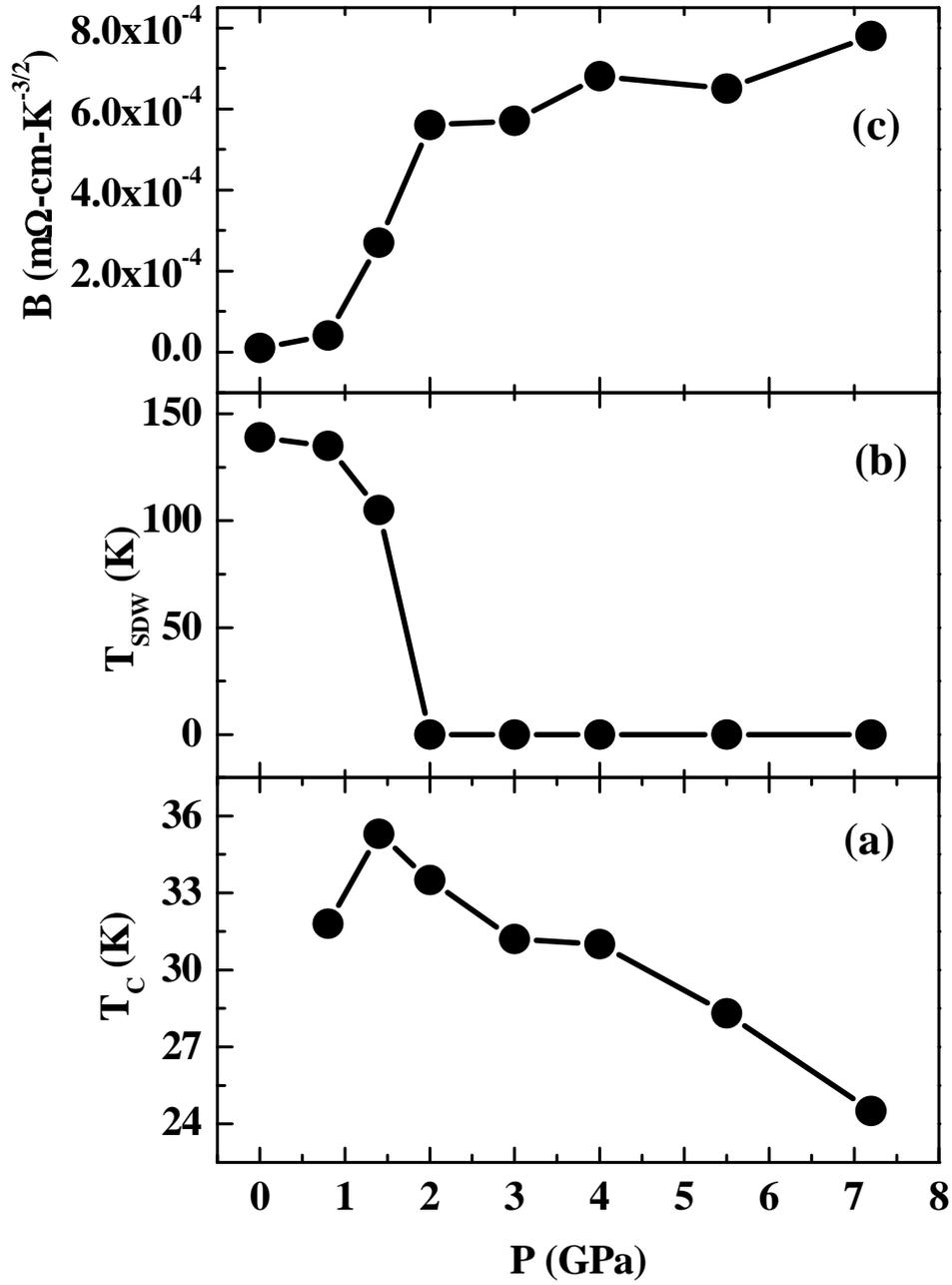